# Superconducting MgB$_2$ films via Precursor Post-Processing Approach

M. Paranthaman, C. Cantoni, H.Y. Zhai, H.M. Christen, T. Aytug,
S. Sathyamurthy, E.D. Specht, J.R. Thompson,
D.H. Lowndes, H. R. Kerchner, and D.K. Christen
Oak Ridge National Laboratory, Oak Ridge, TN 37831-6100

Superconducting MgB$_2$ films with $T_c$ = 38.6 K were prepared using a precursor-deposition, *ex-situ* post-processing approach. Precursor films of boron, ~0.5 mm thick, were deposited onto Al$_2$O$_3$ (102) substrates by e-beam evaporation; a post-anneal at 890°C in the presence of bulk MgB$_2$ and Mg metal produced highly crystalline MgB$_2$ films. X-ray diffraction indicated that the films exhibit some degree of $c$-axis alignment, but are randomly oriented in-plane. Transport current measurements of the superconducting properties show high values of the critical current density and yield an irreversibility line that exceeds that determined by magnetic measurements on bulk polycrystalline materials.

PACS #: 74.70.Ad, 74.76.Db, 74.25.Fy, 74.60.Jg

The discovery of superconductivity at 39 K in $MgB_2$ by Akimitsu et al.[1] has generated intense, worldwide interest. Several groups have rapidly reproduced the results in bulk, polycrystalline $MgB_2$ and are characterizing the material in detail. Bud'ko et al.[2] demonstrated a boron isotope effect in $Mg^{10}B_2$ with an increase of $T_c$ to 40.2 K and concluded that the compound behaves as a phonon-mediated BCS superconductor. Larbalestier et al.[3] demonstrated a strongly linked current flow in hot pressed $MgB_2$ disks. The critical current, determined from magnetization curves as a function of temperature, strongly resembles critical current behavior in low temperature superconductors (LTS) such as $Nb_3Sn$. Magnetically determined critical current densities for porous sintered $MgB_2$ samples were found to be on the order of $10^5$ A/cm$^2$ at 6 K.[4] These results indicate that $MgB_2$ grain boundaries can transmit rather large supercurrents. Dense $MgB_2$ wires were prepared by Canfield et al.[5] by exposing boron filaments to Mg vapor. The resulting wires had a diameter of 160 μm and were 80 % full density. High temporal stability of supercurrents in bulk $MgB_2$ samples have been observed in studies of flux creep, where the creep rate $S = -d\ln(J)/d\ln(t) \leq 0.02$ in fields of 1 Tesla, for temperatures up to $T_c/2$.[6] Compared with high $T_c$ cuprates, the decay rate S is smaller by a factor of 3-10 or more. These results point to possible utility of $MgB_2$ as superconducting wires or tape coatings for conductor applications. According to the Mg-B binary phase diagram,[7] $MgB_2$ decomposes peritectically above 650 ºC and has no exposed liquid-solidus surface. This behavior will restrict approaches to single crystal growth. Pulsed laser deposition techniques have already been used to grow $MgB_2$ films on various substrates followed by an ex-situ[8-10] or in-situ[11] anneal. The $T_c$ varies from 12 to 39 K in these reports. Here, we report our successful demonstration of the growth of superconducting $MgB_2$ films using electron beam evaporated B precursor films followed by appropriate post-annealing. We also report results of transport property measurements on these ex-situ grown $MgB_2$ films.

Electron beam evaporation was used to deposit B films directly on $Al_2O_3$ (102) single crystal substrates with dimensions of 0.35 cm x 1.2 cm at room temperature at a base pressure of 1 x $10^{-6}$ Torr. The deposition rate for B was 10-12 Å/sec with the operating pressure of $10^{-5}$ Torr, and the final thickness was 5000 Å to 6000 Å. The shiny amorphous B films were sandwiched between cold-pressed $MgB_2$ pellets, along with excess Mg turnings, and packed inside a crimped Ta cylinder. The polycrystalline $MgB_2$ powders were prepared by a solid-state reaction of stoichiometric Mg turnings and B in a sealed Ta cylinder at 890°C for 2 hours. The Ta cylinder containing the precursor film was then introduced into a quartz tube, evacuated to 1 x $10^{-5}$ Torr, and sealed. The sealed quartz capsule was placed inside a box furnace, where the samples were heated rapidly to 600ºC, and maintained there for 5 minutes. Then the furnace temperature was rapidly increased to 890ºC, held at 890ºC for 10-20 minutes, and then furnace-cooled to room temperature. The as-formed purplish gray film had a very low two-probe resistance of < 1 Ohm. The $MgB_2$ films were analyzed by X-ray diffraction. Hitachi S-4100 field emission scanning electron microscope was used to take images with a beam voltage of 15 kV. The thickness of the films was determined by Alpha Step profilometer scans. $T_c$ and $J_c$ were measured using a standard four-probe method, at a criterion of 1 $\mu$V/cm to define $J_c$. During the $J_c$ measurements, a magnetic field (H) was applied perpendicular to the film, and the irreversibility field was defined according to the emerging voltage-current power-law characteristic, $V \propto I^2$.

Typical θ-2θ scans for $MgB_2$ films on $Al_2O_3$ single crystal substrates are compared with polycrystalline $MgB_2$ powders in Figure 1. The strong $MgB_2$ (001) and (002) signal revealed the presence of a c-axis aligned film. The full width at half maximum (FWHM) value for $MgB_2$ (002) omega scans was 4.45 °. However, the $MgB_2$ (101) pole figure indicated that the film has random in-plane texture. The observation of random in-plane texture in $MgB_2$ films could be due to the initial reaction of Mg vapors with B films at the free surface of the film. This causes the bulk crystallization of $MgB_2$ to occur rather than the epitaxial nucleation of $MgB_2$ at the substrate/film interface. Figure 2 shows the microstructure of $MgB_2$ films determined by SEM. The films have a dense microstructure with large grains present. Both c-axis and random grains are apparently present. The temperature-dependent resistivity near $T_c$ is shown in the inset of Fig. 1 for a 5700 Å thick $MgB_2$ film, which had a room temperature resistivity near 12 μOhm-cm. The $MgB_2$ films had a sharp $T_c$ (zero resistance) of 38.0 K with a $\Delta T_c$ of 0.3 K and a ratio of the room temperature resistivity to the residual resistivity above $T_c$ of about 2. The resistivity decreased linearly with temperature indicating that the $MgB_2$ film is metallic. We have also produced $MgB_2$ films with a high $T_c$ (zero resistance) of 38.6 K with a $T_c$ onset of 39.0 K. At present, $MgB_2$ film with a $T_c$ of 32.5 K was patterned to 250 μm bridge to perform the $J_c$ measurements. The field dependent transport critical current density, $J_c$ for a 6300 Å thick $MgB_2$ film is shown in Figure 3. A transport $J_c$ of 2 x $10^6$ A/cm$^2$ at 20 K was obtained in zero field, while at 1 Tesla, $J_c$ decreased to 2.5 x $10^5$ A/cm$^2$. Figure 2 inset shows the temperature dependence $J_c$ in self field. At 5 K, a transport $J_c$ of 4 x $10^6$ A/cm$^2$ at self-field was obtained, although this film had a $T_c$ zero of 32.5 K; this level of current conduction is very comparable to that observed magnetically in isolated grains of $MgB_2$, for the same conditions. The temperature dependence of irreversibility field $B_{irr}$ obtained from transport measurements on $MgB_2$ films, is shown in Figure 4. These data are compared with those obtained from magnetization measurements[6] on polycrystalline sintered $MgB_2$ pellets. These data are quite comparable down to 26 K. At low temperatures, there is some enhancement in the film's $B_{irr}$, indicating that there is an improvement in flux pinning properties in these $MgB_2$ films. The temperature dependence is well described by $B_{irr}$ $\alpha$ $(1-T/T_c)^{3/2}$ which is depicted by the solid line fitted to the experimental data. For comparison, the $H_{c2}$ data[5] for $MgB_2$ are also plotted (solid line). Further improvements in $B_{irr}$ may be possible for epitaxial films and efforts are underway to produce epitaxial films on lattice matched substrates using either in-situ or ex-situ methods.

In summary, we have prepared superconducting $MgB_2$ films with a sharp $T_c$ of 38.6 K on $Al_2O_3$ single crystal substrates using electron beam evaporated B films followed by post-annealing. Detailed X-ray diffraction studies indicate that the film is polycrystalline with some degree of c-axis texture. A transport $J_c$ of 2 x $10^6$ A/cm$^2$ was obtained on $MgB_2$ films at 20 K. The irreversibility field, $B_{irr}$ obtained from the transport measurements on $MgB_2$ films indicate that there may be some improvement in flux pinning at lower temperatures.

**ACKNOWLEDGEMENTS**

Thanks are due to Pam Fleming for evaporating B films. This work was supported by the U.S. Department of Energy, Division of Materials Sciences, Office of Science, Office of Power Technologies-Superconductivity Program, Office of Energy Efficiency and

**FIGURE CAPTIONS**

Figure 1    A typical $\theta$-$2\theta$ scans for (a) $MgB_2$ film on $Al_2O_3$ (102) substrate, and (b) powdered $MgB_2$ material. The $MgB_2$ film has a preferred c-axis orientation. Inset is the expanded version of the resistivity plot. $MgB_2$ films had a $T_c$ (zero resistance) of 38.0 K and a $T_c$ onset of 38.3 K.

Figure 2    SEM micrograph for 6300 Å thick $MgB_2$ film on $Al_2O_3$ substrate, indicating the presence of a granular microstructure.

Figure 3    The field dependence of the transport critical current density, $J_c$ for a 6300 Å thick $MgB_2$ film on $Al_2O_3$ substrate at 20 K. Inset is the temperature dependence $J_c$ for the same film at H = 0.

Figure 4    The temperature dependence of the irreversibility field, $B_{irr}$ obtained from a transport measurement for $MgB_2$ film on $Al_2O_3$ substrate (closed triangles) are compared with those obtained from the polycrystalline 61% dense $MgB_2$ pellets (closed circles) (obtained from reference 6) and $H_{c2}$ data for bulk $Mg^{10}B_2$ sample (solid line) (obtained from reference 4)

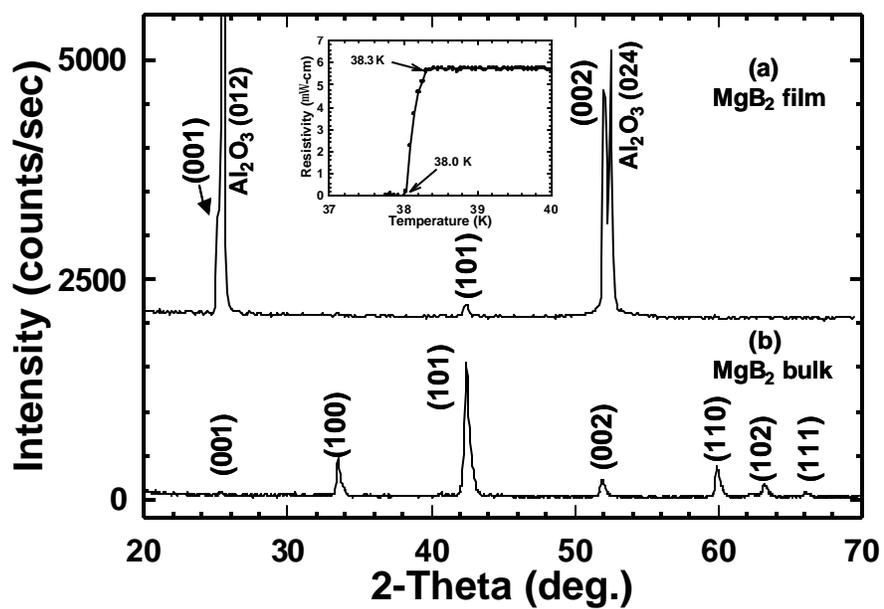

Figure 1 Paranthaman et al

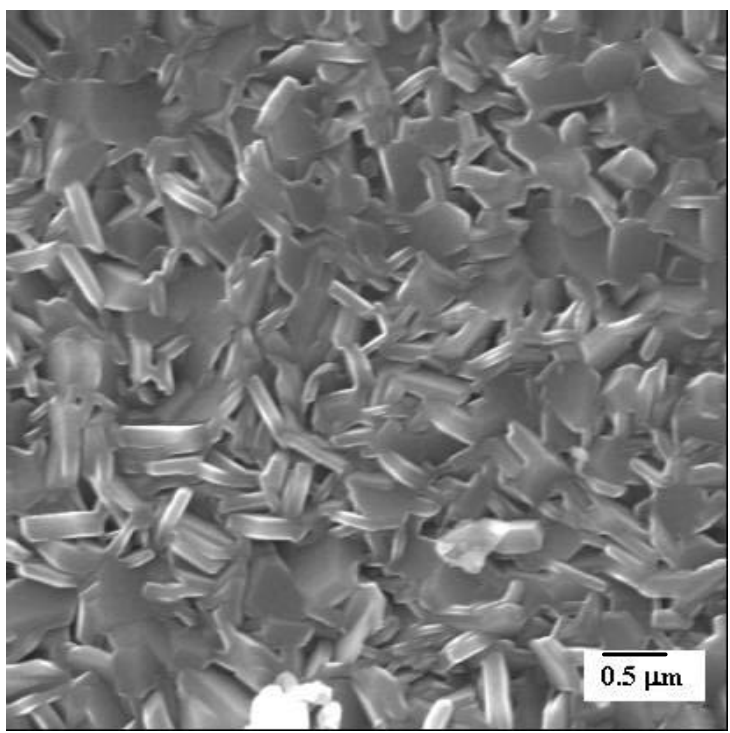

Figure 2 Paranthaman et al

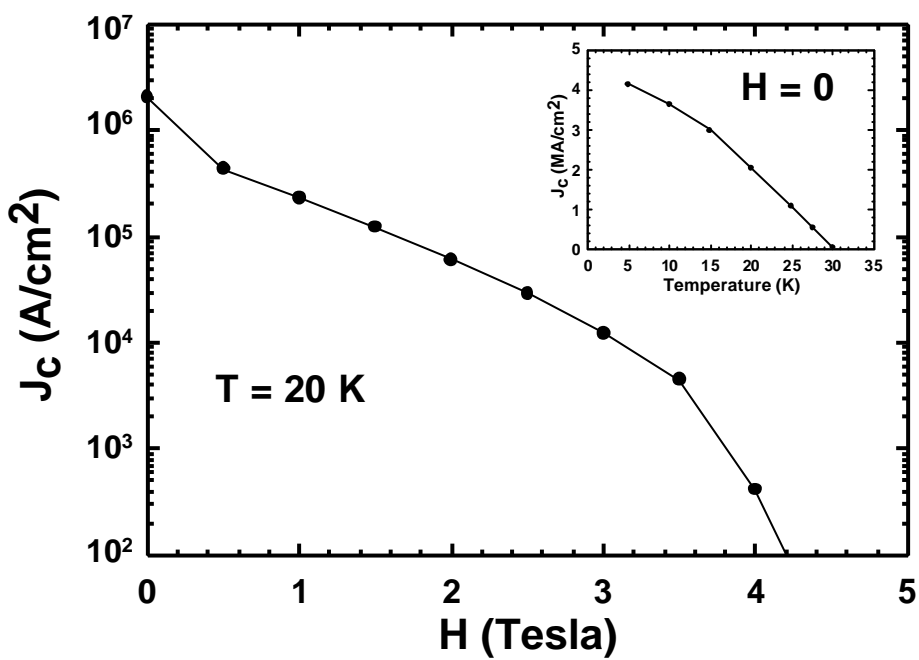

Figure 3 Paranthaman et al

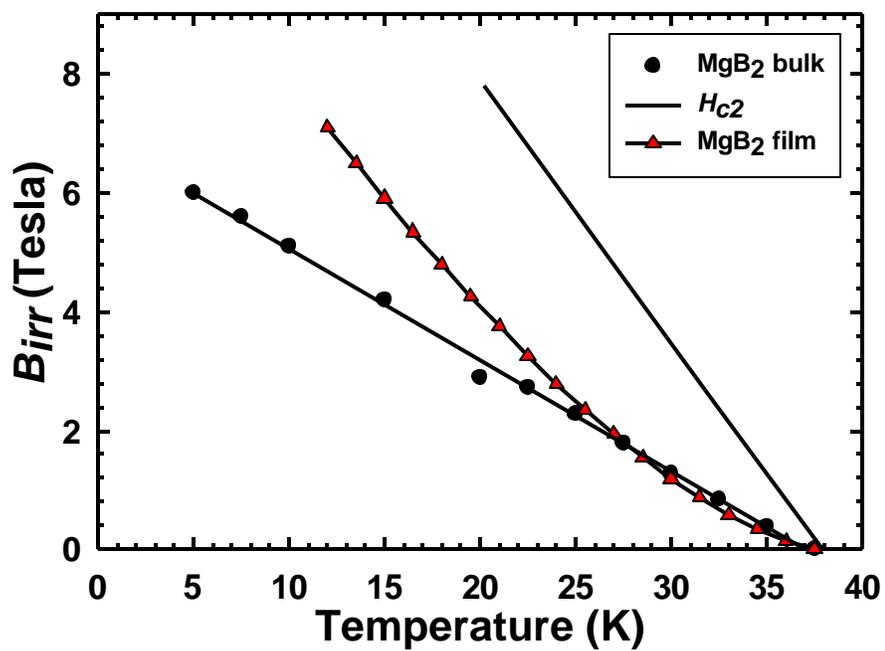

Figure 4 Paranthaman et al